\documentclass[12pt]{article}
\usepackage{graphicx,cite}
\newcommand{\PSbox}[3]{\mbox{\rule{0in}{#3}\includegraphics{#1}\hspace{#2}}}
\newcommand{\beq}{\begin{eqnarray}}
\newcommand{\eeq}{\end{eqnarray}}
\def\be{\begin{equation}}
\def\ee{\end{equation}}
\def\ba{\begin{eqnarray}}
\def\ea{\end{eqnarray}}

\newcommand{ \sla }[1]{\setbox0=\hbox{$#1$}         
   \dimen0=\wd0                                     
   \setbox1=\hbox{/} \dimen1=\wd1                   
   \ifdim\dimen0>\dimen1                            
      \rlap{\hbox to \dimen0{\hfil/\hfil}}          
      #1                                            
   \else                                            
      \rlap{\hbox to \dimen1{\hfil$#1$\hfil}}       
      /                                             
   \fi}                                             %

\def\lesssim{\mathrel{\mathpalette\vereq<}}

\makeatletter
\def\vereq#1#2{\lower3pt\vbox{\baselineskip1.5pt \lineskip1.5pt
\ialign{$\m@th#1\hfill##\hfil$\crcr#2\crcr\sim\crcr}}}
\makeatother

\begin{document}

\begin{titlepage}
\noindent
\begin{flushright}
SU-ITP-01/48\\
{\tt hep-ph/0111311}\\
\end{flushright}

\vskip0.5cm
\begin{center}
{\LARGE{\bf Dimming Supernovae without Cosmic}} \\
\vskip .05in
{\LARGE{\bf Acceleration}} \\

\end{center}
\vskip0.5cm
\begin{center}
{\bf Csaba Cs\'aki$^{a,}$\footnote{Address after December 20, 2001: 
Newman Laboratory of Physics, Cornell University, Ithaca, NY 14853.
E-mail: {\tt csaki@mail.lns.cornell.edu}},
Nemanja Kaloper$^{b}$ and
John Terning$^{a}$}
\end{center}

\vskip 10pt

\begin{center}
$^a${\em Theoretical Division T-8, Los Alamos National Laboratory,
Los Alamos, NM 87545}\\

\vskip 0.1in

$^b${\em Department of Physics, Stanford University,
Stanford CA 94305-4060 }\\

\vskip 0.1in

\vskip 0.1in
{\tt  csaki@lanl.gov, kaloper@stanford.edu, terning@lanl.gov}

\end{center}

\vskip .25in
\begin{abstract}
We present a simple model where photons propagating
in extra-galactic magnetic fields can oscillate into very light
axions.
The oscillations may convert some of the photons departing a distant supernova
into axions, making the supernova appear dimmer
and hence more distant than it really is. Averaging over different
configurations of the magnetic field we find that the dimming
saturates at about 1/3 of the light from the supernovae at very large
redshifts. This results in a luminosity-distance vs.~redshift curve almost
indistinguishable from that
produced by the accelerating Universe, if the
axion mass and coupling scale are $m \sim 10^{-16}$ eV,
$M\sim 4\cdot 10^{11}$ GeV. This phenomenon may be
an alternative to the accelerating Universe for explaining
supernova observations.
\end{abstract}

\end{titlepage}

\setcounter{equation}{0} \setcounter{footnote}{0}

Current observations of supernovae (SNe) at redshifts
$0.3 \lesssim z \lesssim 1.7$
reveal that they are fainter than expected from the luminosity-redshift
relationship in a decelerating Universe~\cite{supnov}.
On the other hand, the large scale structure and CMBR
observations suggest that the Universe is spatially flat,
with the matter density about $30\%$ of the critical
density~\cite{krtur,ostrstei,revcospar,boom,knox}.
It is therefore usually inferred that the Universe must have become
dominated by a dark energy component,
which comprises about $70\%$ of the critical energy density,
and has the equation of state
$p/\rho \lesssim -2/3$, implying that our Universe
would be accelerating at present. The dark energy
component could either be a small cosmological constant
or a time-dependent quintessence energy~\cite{q}.
Neither possibility is elegant from the current vantage point of
fundamental theory because of unnaturally small numbers needed to
fit the data: the present value of the energy density, $\rho_c
\sim 10^{-12}$ eV$^4$, and, in the case of quintessence, the tiny
mass smaller than the current Hubble parameter $m_Q < H_0 \sim
10^{-33}$ eV and sub-gravitational couplings to visible matter to
satisfy fifth-force constraints~\cite{qbounds}. Further,
describing an eternally accelerating Universe with future horizons
is at present viewed as a conceptual challenge for string theory
and more generally any theory of quantum gravity~\cite{banks,hkl,fkmp}.

Because the SNe observations probe length scales
$l \sim H^{-1}_0 \sim {\rm few} \times 10^3$ Mpc which are
inaccessible to
any particle physics experiments, it is natural to
consider alternative explanations to the supernova data without
cosmological dark energy. A simple such alternative is
that light emitted by a distant
supernova encounters an obstacle en route to us
and gets partially absorbed~\cite{dust}.\footnote{Other
possibilities are that SNe may evolve with time \protect\cite{evolution},
or that there are more than four space-time dimensions\protect\cite{Gia}.}
However any mechanism must be very achromatic because
the light from
the SNe appears to be dimmed
independently of frequency. This would seem to rule out a medium
of matter particles, which can absorb light in the optical spectrum
but will re-emit it in the IR, affecting the CMBR in unacceptable ways.

In this paper we consider a model where the
dimming of SNe is based on flavor oscillations.
Flavor oscillations occur whenever there are several degrees of freedom
whose interaction eigenstates do not coincide with the
propagation eigenstates. Such particles can
turn into other particles simply by evolution and
evade detection. We will consider a model with
an axion with a mass $m \sim 10^{-16}$ eV, much smaller than the
usual~\cite{axion} QCD axion mass scale, $10^{-5}$ eV $\lesssim m_{QCD}^{ax}
\lesssim 10^{-1}$ eV, but exponentially
larger than the quintessence mass $m_Q \lesssim 10^{-33}$ eV.
This axion couples to electromagnetism through the
usual term $(a/4M) \tilde F F$, which leads to energy-dependent
mixing of the photon and the axion in the presence of an external
magnetic field $B$ \cite{Raffelt}. Hence light traveling in
inter-galactic magnetic fields can in part turn into axions,
and evade detection on Earth. A source would then appear
fainter even if the Universe is not accelerating. To satisfy
other cosmological constraints we assume that the Universe is
presently dominated by some form of uniform dark energy which
does not clump, but need not lead
to cosmological acceleration, e.g.~with equation of state
$p/\rho =-1/3$.

We find that contrary to the familiar
example of neutrino oscillations, in our model both the flavor mixing
and the oscillation length of photons in the optical range are
insensitive to energy, and so our axion will induce strongly achromatic
oscillations of optical photons. On the other hand
the small axion mass  $m \sim 10^{-16} {\rm eV}$ insures
that the photon-axion oscillations leave CMBR essentially
unaffected. Our task here is to find
a small correction to the luminosity-redshift relation induced by
the oscillations. Because the SNe observations at redshifts
$ z \sim 0.5$ can be explained by a $10 \%-15 \%$ total increase in the
distance relative to a matter dominated universe,
the total attenuation of SNe light should be about $30 \%$
between the cosmic string-dominated geometry and
flavor oscillations. Hence we need to account for about a $20\%$ of decrease
in the luminosity by the flavor oscillations.

The axion-photon coupling is
\be
{\cal L}_{int}=
\frac{a}{M} \vec E \cdot \vec B \, ,
\label{coupling}
\ee
where the scale $M$ characterizes the strength of the axion-photon
interactions.
This induces a mixing between the photon
and the axion~\cite{Raffelt,GGG} in the presence of a background magnetic
field $\vec B$
(as exists in our Universe~\cite{kronberg}).  Indeed,
working in the Coulomb gauge at distances short compared
to the size of a coherent magnetic domain $L_{dom}$,
we see that the photon with electric field orthogonal
to $\vec B$ remains unaffected by mixing.
The polarization whose electric field is parallel to $\vec B$
mixes with the axion. The field equations are,
after rotating the coordinate axes such that the propagation is along the
$y$-direction,
\be
\Bigl\{ \frac{d^2}{dy^2}
+ {\cal E}^2 - \pmatrix{ 0 & i {\cal E} \frac{B}{M_{L}} \cr
-i {\cal E} \frac{B}{M_{L}} & m^2 \cr} \Bigr\}
\pmatrix{ |\gamma\rangle  \cr |a\rangle \cr} = 0
\label{frice}
\ee
where we Fourier-transformed the fields to the
energy picture ${\cal E}$ and introduced the
state-vectors $|\gamma\rangle$ and $|a\rangle$ for the photon and
the axion. Here $B = \langle \vec e \cdot \vec B\rangle \sim | \vec B|$
is the averaged projection of the extra-galactic magnetic field
on the photon polarization $\vec e$. We will assume that the
averaged value of $\vec B$ is close to its observed upper limit,
and take for the magnetic field amplitude
$|\vec B| \sim {\rm few} \cdot 10^{-9}$ G \cite{kronberg,FL}.
Therefore its energy
density is $\vec B^2 \sim c H_0^2 M^2_{Pl}$, where $c\sim {\rm few}
\cdot 10^{-11}$
and the Hubble parameter is $H_0 \sim 10^{-33}$ eV. The
magnetic fields we will be considering are sufficiently small that
we can safely ignore the Euler-Heisenberg
effect~\cite{Raffelt,Heisenberg}.

We can now define the propagation eigenstates by diagonalizing the mixing
matrix in Eq.~(\ref{frice}), which is, using $B/M = \mu$,
\be
{\cal M}^2 = \pmatrix{ 0 & i {\cal E} \mu \cr
-i {\cal E} \mu & m^2 \cr} \, .
\label{massm}
\ee
This matrix is the analogue of the usual see-saw matrix for
neutrinos, with the only difference that the off-diagonal terms
are imaginary and complex-conjugates of each other. This
is because the mixing arises from the derivative terms in the
field equations rather than the potential terms.
Defining the propagation eigenstates $|\lambda_-\rangle$ and
$|\lambda_+\rangle$
which diagonalize the matrix (\ref{massm}), whose
eigenvalues are $\lambda_\mp = \frac{m^2}{2} \mp \sqrt{\frac{m^4}{4}
+ \mu^2 {\cal E}^2} $,
we can solve the Schr\" odinger equation (\ref{frice}). The
solutions describing particles emitted by a supernova
at a distance $y_0>0$ and propagating towards us at $y=0$ are
\ba
|\gamma\rangle &=& \frac{ \mu{\cal E}}{\sqrt{\lambda_-^2 +
\mu^2{\cal E}^2}} |\lambda_-\rangle \ e^{-i[{\cal E}t + p_1 (y-y_0)]}
+ \frac{i \mu{\cal E}}{ \sqrt{\lambda_+^2 +
\mu^2{\cal E}^2}} |\lambda_+\rangle \ e^{-i[{\cal E}t + p_2 (y-y_0)]}
\,~ ,
\nonumber \\
|a\rangle &=&\frac{-i \lambda_-}{\sqrt{\lambda_-^2 +
\mu^2{\cal E}^2}} |\lambda_-\rangle \ e^{-i[{\cal E}t + p_1
(y-y_0)]} +
\frac{ \lambda_+}{ \sqrt{\lambda_+^2 +
\mu^2{\cal E}^2}} |\lambda_+\rangle \ e^{-i[{\cal E}t + p_2 (y-y_0)]}
\,~ ,
\label{solns}
\ea
where $p_k = \sqrt{{\cal E}^2 - \lambda_k}$.
It is now clear that as the photon propagates, it mixes with
the axion by an amount depending on the energy of the particle.
In the limit ${\cal E}^2 \gg \lambda_{i} > m^2$, which covers
all of the applications of interest to us, the mixing angle is
\be
\sin \theta = \frac{\mu {\cal E}}{\sqrt{\lambda^2_+ + \mu^2 {\cal
E}^2}} \, ,
\label{mixang}
\ee
the photon survival probability
$P_{\gamma \rightarrow \gamma} = |\langle \gamma(y_0)|\gamma(y)\rangle|^2$
is
\be
P_{\gamma \rightarrow \gamma} = 1 - \frac{4 \mu^2 {\cal E}^2}{m^4+4\mu^2
{\cal E}^2}
\sin^2\left[  \frac{\sqrt{m^4 + 4 \mu^2 {\cal E}^2}}{4{\cal E}} \Delta y\right]
\, ,
\label{prob}
\ee
and the oscillation length is
\be
L_O = \frac{4\pi {\cal E}}{\sqrt{m^4 + 4 \mu^2 {\cal E}^2}} \, .
\label{oscl}
\ee
In the limit ${\cal E} \gg m^2/\mu$, the mixing is
maximal, while the oscillation
length is completely independent of the
photon energy: $
\sin \theta \sim  \frac{1}{\sqrt{2}}$, $L_O \sim \frac{2\pi}{\mu}$.
Thus high-energy photons (including optical frequencies
${\cal E} \sim 10$ eV as we will see) oscillate achromatically.

On the other hand, in the low energy limit ${\cal E} \ll m^2/\mu$,
the mixing is small, and the
oscillation length is sensitive to energy:
$\sin \theta \sim \frac{\mu {\cal E}}{m^2} $,
$L_O \sim \frac{4\pi {\cal E}}{m^2}$.
The oscillations are very dispersive,
due to the energy-dependence of both the mixing angle
and the oscillation length.
But the probability to find axions $P_{\gamma \rightarrow a} =
1 - P_{\gamma \rightarrow \gamma}$ is small, bounded from above by
$P_{\gamma \rightarrow a} < \sin^2(2 \theta) \le 4 \mu^2 {\cal
E}^2/m^4$.

In our Universe the magnetic field is not uniform.
Assuming that a typical
domain size for the extra-galactic magnetic field
is $L_{dom} \sim $ Mpc \cite{kronberg,FL},
it is straightforward to numerically solve
for the quantum mechanical
evolution of unpolarized light in such magnetic domains
with uncorrelated field directions.
An analytic calculation shows that in the case of maximal mixing,
with $\cos (\mu L_{dom}) > -1/3$, the survival probability is
monotonically decreasing:
\be
P_{\gamma \rightarrow \gamma} = \frac{2}{3}+\frac{1}{3}
e^{- \Delta y/L_{\rm decay}}
\ee
where the inverse decay length is given by
\be
L_{\rm decay}= \frac{L_{\rm dom}}{ \ln \left(
\frac{4}{1+3\cos (\mu L_{\rm dom})}\right)}~.
\ee
For $\mu L_{\rm dom} \ll 1$ this reduces to
\be
L_{\rm decay}= \frac{8}{3 \mu^2 L_{\rm dom}}~.
\ee
Thus we see that with a random magnetic field the problem becomes essentially
classical and after the traversal of many magnetic domains the system is
equilibrated
between the two photon polarizations and the axion.  This leads to the
generic prediction that on average
one-third of all photons converts to axions after large traversed
distances.

We can now estimate the axion mass
and coupling needed to reproduce SN observations.
To take the oscillations into account, in the luminosity-distance v.s.
redshift formula we should replace the absolute luminosity ${\cal L}$
by an effective one:
\be
{\cal L}_{eff} = {\cal L} ~P_{\gamma \rightarrow \gamma} \, .
\ee
The optical photons must oscillate independently of their
frequency. For them,
the oscillations should reduce the flux of incoming
photons by about $20 \%$ for SNe at $z \sim 0.5$. This requires
$L_{\rm dec} \lesssim H^{-1}_0/2$. Thus the mass scale $M$ for this
should be $M \sim 4\cdot 10^{11}$ GeV. Note, that this is above the
experimental exclusion limit for $M$. The experimental bound on
$M$ quoted by the PDG~\cite{PDG}
is $M \geq 1.7\cdot 10^{10}$ GeV~\cite{Raffeltreview}. However, for
ultralight axions there is~\cite{Raffeltreview,SN1987A}
a more stringent (though also more model
dependent) limit from SN1987A given by $M \geq 10^{11}$ GeV, which is still
lower than the value required here.

If the microwave photons
had fluctuated a lot in the extra-galactic magnetic field, their
anisotropy would be very large due to the
variations in the magnetic field.
To avoid affecting the small primordial CMBR anisotropy, $\Delta T/T \sim
10^{-5}$, the axion mass should be
large enough for the mixing between microwave photons
and the axion to be small. In this limit, we can ignore the
averaging over many random magnetic domains and simply
treat each domain as a source of CMBR fluctuation.
The photon-axion mixing and the oscillation length
in that case are given by $\sin \theta \sim \frac{\mu {\cal E}}{m^2} $,
$L_O \sim \frac{4\pi {\cal E}}{m^2}$. The
disturbances of CMBR are controlled by the transition
probability into axions $P_{\gamma \rightarrow a} \le 4 \frac{B^2 {\cal
E}^2}{m^4 M^2}$, which using the explicit expression for $B$ is
\be
P_{\gamma \rightarrow a} \le 4\cdot 10^{-11}  \frac{M_{Pl}^2 H_0^2 {\cal
E}^2}{M^2 m^4} \, .
\ee
For microwave photons ${\cal E} \sim 10^{-4}$ eV, and so
$P_{\gamma \rightarrow a} \le 2.5 \cdot 10^{-70} { ({\rm eV})^4}/{m^4} $.
Therefore for $m \sim {\rm few} \times 10^{-16}$ eV we find
$P_{\gamma \rightarrow a} \le 10^{-7}$, which is smaller than the
observed temperature anisotropy. For this mass scale, the
oscillation length of microwave photons is $L_O \sim 10^{-4}
H_0^{-1}$, which is of order of the coherence length
of magnetic domains $L_{dom}$ and so a lot shorter than
the horizon size. This is
harmless since the oscillation amplitude is so small.
Thus we see that if the axion scales are
\be
m \sim 10^{-16} {\rm eV} \, ,
~~~~~~~~~ M \sim 4\cdot 10^{11} {\rm GeV} \, ,
\label{finscale}
\ee
the mixing  could produce the desired effect of
reducing the flux of light from SNe while leaving the primordial CMBR
anisotropy unaffected.
We stress here that while at early times the CMBR photons were
much more energetic there were no sizeable extra-galactic
magnetic fields yet, since their origin is
likely tied to structure formation \cite{egmf}. Hence
we can get a rough estimate of the influence of our effect on CMBR
using their current energy scale.
Having determined the scales, we can check that the
approximations we have been using are appropriate
for optical and microwave photons, respectively. In the former
case, the mixing angle and the oscillation length receive
energy-dependent corrections $\propto \frac{m^4}{\mu^2 {\cal
E}_{\rm optical}^2} \sim 10^{-5}$, while in the latter case the
energy-dependent corrections are $\propto \frac{\mu^2 {\cal
E}_{\rm CMB}^2}{m^4} \sim 10^{-6}$, confirming the validity of
our approximations.

\begin{figure}
\PSbox{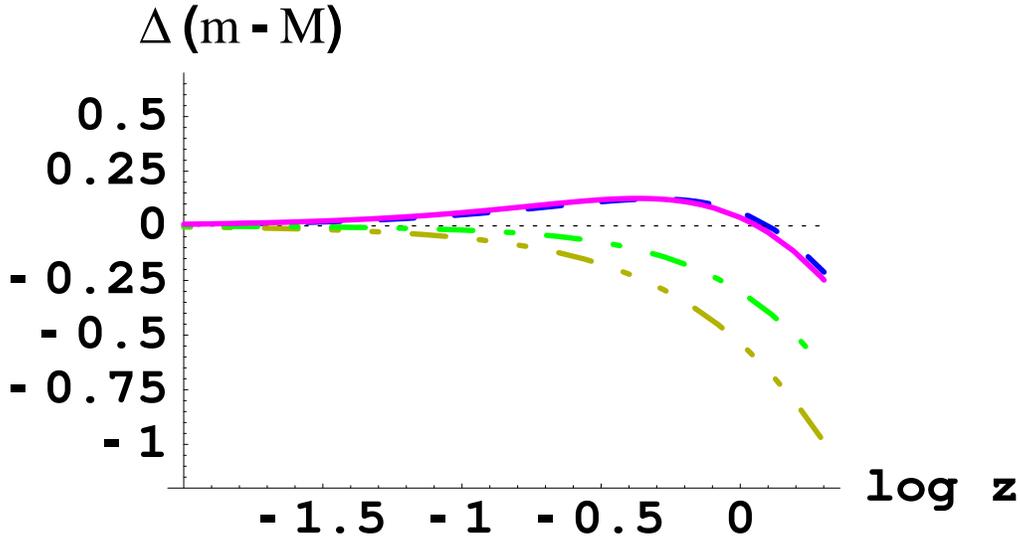 hscale=70 vscale=70 hoffset=0
voffset=-120}{7cm}{6cm}
\caption{{\small The luminosity-distance vs. redshift curve
for several models, relative to the curve with $\Omega_{tot}=0$
 (dotted horizontal
line). The dashed curve is a best fit to the supernova data
assuming the Universe is accelerating
($\Omega_m=0.3$, $\Omega_\Lambda=0.7$); the solid line is the
oscillation model with $\Omega_m=0.3$, $\Omega_S=0.7$,
$M =4\cdot 10^{11}$ GeV, $m=10^{-16}$ eV;
the dot-dashed line is $\Omega_m=0.3$, $\Omega_S=0.7$
with no oscillations, and the dot-dot-dashed line is for $\Omega_m=1$
again with no oscillations.}
\label{fig:lum}}
\end{figure}

To compare our model with observations, we assume that the constraint
on the total energy density of the Universe $\Omega_{tot} \simeq 1$ is
satisfied because the Universe contains some form of dark energy which
does not clump, but it need not lead to cosmological acceleration.
A simple example is dark energy with the equation
of state $w=p/\rho =-1/3$ and energy density
$\Omega_S=0.7$, which could originate from a
network of frustrated strings with small mass per unit length.
Note that because the scale factor $a$ of the universe obeys
$\ddot a/a = - \frac{4\pi}{3M^2_{PL}} (\rho_{total} + 3p_{total})$,
and assuming $\Omega_m = 0.3$ and $\Omega_{dark} = 0.7$,
then as long as the ratio $w = p/\rho$ is greater than $-1/2.1
\simeq -0.48$
the Universe would presently not be
accelerating. These forms of dark energy do not
appear to be
excluded either by the position of the first acoustic peak in the CMBR
measurements~\cite{Turner1} or by combined CMBR+large scale
structure fits~\cite{PTW}. In Fig.~\ref{fig:lum} we have
plotted the typical prediction of the oscillation model in a
spatially flat Universe with $\Omega_m=0.3$ and $\Omega_S=0.7$
against the best fit model for the accelerating Universe with a cosmological
constant ($\Omega_m=0.3$ and $\Omega_\Lambda=0.7$). The two
curves are practically indistinguishable. We note  that
the oscillation model predicts limited attenuation of the SN luminosities,
unlike some other alternatives to the accelerating Universe. The total
attenuation is limited to about $1/3$ of the initial luminosity,
as we have explained above. Since for larger values of $z$ the Universe
becomes matter dominated, and the disappearance of photons is saturated
in the oscillation model, the two curves will continue lying on top of
each other for higher values of $z$. Thus simply finding higher $z$
supernovae~\cite{TurnerRiess} will not distinguish between the two models.
The main difference between the two
is that the curve for the oscillation model is an averaged curve,
with relatively large standard deviations. Therefore it may be much easier
to explain outlying events than in the case of the accelerating Universe.

Let us now consider photons which may pass
through the magnetic field of
a galaxy, or just skim it.
The galactic magnetic fields are much stronger
than the extra-galactic ones, $B_G \sim \mu {\rm G} \sim 10^3 B$.
However, the density of baryons (and therefore also of electrons)
is large enough in such regions that refraction has to be
taken into account, which introduces a diagonal element ${\cal M}_{11}$
for the photon in (\ref{massm}).\footnote{We thank Georg Raffelt for pointing
out this effect.} A simple estimate~\cite{Raffelt} for this term gives
${\cal M}_{11} \sim 10^{-23} ({\rm eV})^2$ for 10 eV photons
traveling within a galaxy, while the
off-diagonal terms are of the order
$10^{-27} ({\rm eV})^2$. Therefore this term will dominate the mixing
matrix, and the oscillations will be highly suppressed
while passing through the magnetic
field of a galaxy. However, since there is no evidence for the
presence of gas uniformly distributed between clusters, this effect
is likely negligible for most of 
extra-galactic space. This is because a simple estimate shows, that 
even assuming the worst-case scenario where all matter is uniformly
distributed and totally ionized, ${\cal M}_{11}$ would be $\sim 10^{-29}
 ({\rm eV})^2$ for $\Omega_{baryon}=0.05$, which would somewhat suppress
the mixing. However since matter is not uniformly 
distributed and definitely not ionized in the inter-cluster voids 
(which make up most of space) this effect should be negligible for our results.

While it is natural to wonder if there are laboratory
constraints on our mechanism, a simple order of magnitude
estimate shows that it would be difficult to observe in a lab.
Since we have assumed that the extra-galactic magnetic field is
$\sim 10^{-13}$ T, for a uniform
magnetic field about $10^{14}$ times larger the corresponding
oscillation length would be $L_O/10^{14} \sim 6\cdot 10^{12}$ cm,
which is about a thousand times the circumference of the Earth.
The current direct experimental bounds~\cite{Moriyama}
quoted by the PDG~\cite{PDG} on
the coupling of an axion-like particle (with a mass less than 0.03
eV) to $\vec E \cdot \vec B$ is $M > 1.6 \times 10^{9}$ GeV.

\begin{figure}
\PSbox{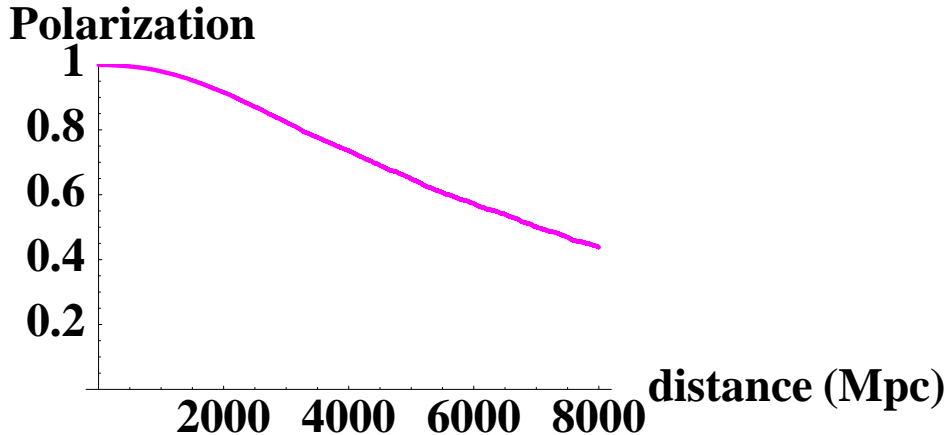 hscale=65 vscale=65 hoffset=30
voffset=-140}{7cm}{6cm}
\caption{{\small The effect of the oscillation on the polarization
of the photons. We assume that the photon emitted is totally polarized,
and show how much polarization remains as a function of physical
distance traveled.}
\label{fig:pol}}
\end{figure}

Another question is whether the oscillations
may cause any observable
polarization effects on the light arriving from the SNe.
If the orientation of the extra-galactic magnetic field were constant, and the
field perfectly homogeneous, light from the SNe would be
partially plane-polarized. However, since the coherence length of
the extra-galactic magnetic field is of order $\sim$ Mpc,
the direction of the magnetic field is effectively random,
and thus no strong polarization effects are expected for faraway SNe.
Rather, the converse effect of depolarizing incoming light is
more important, since the oscillations in a random
magnetic field may deplete existing photon beam polarizations.
Because there are distant sources which are
partially polarized, with the polarization direction correlated with
the shape of the source, it is important to show
that the photon-axion mixing
does not completely depolarize light from a polarized source.
The observed polarization as a function of distance is shown in
Fig.~\ref{fig:pol},
where we see that the polarization decrease is rather slow.
This should be expected because the degradation of polarization
occurs after an axion produced by a polarized photon conversion
regenerates a photon of a different polarization, after the
orientation of the $\vec B$ has changed. This is a second-order
effect, and so polarization is depleted more slowly than intensity.
As a result the existing measurements of polarized optical photons from distant
sources can be accommodated in this model.

An axion with the scales
which are needed for our model can for example
arise from the spontaneous breaking of an
axial lepton number symmetry.  Suppose that it
couples to the electroweak gauge theory in the standard way.
Specifically it would
couple to the electromagnetic field
like the QCD axion~\cite{axion}.
Then the mass scale $M$ would be
related to the scale of the spontaneous breaking of axial symmetry $f_a$ by
\be
M = \frac{8\pi}{\alpha} \frac{f_a}{\xi} \, ,
\label{cp}
\ee
where $\alpha=g^2/4\pi$ is related to the gauge coupling constant,
and $\xi$ is a dimensionless number depending on the precise couplings
to fermions.
We will take $\alpha \sim 1/30$ and $\xi = {\cal O}(1)$ in what follows
for simplicity. Hence $f_a \sim 10^{-3} M$.
In perturbation theory the shift symmetry $a
\rightarrow a + c$ protects the axion from acquiring a mass term
(more generally any potential). This symmetry is broken
by nonperturbative effects induced by
instantons, which give rise to the axion potential~\cite{axion}.
Because by assumption our axion couples to
electroweak gauge fields, a possibility to generate the potential
is via the electroweak instantons. In particular
the axion potential will be of the form
\be
V(a) = \Lambda^4 \Bigl[ 1 + \cos(\frac{a}{f_a}) \Bigr] \, .
\label{axpot}
\ee
For example, in a particular
supersymmetric (SUSY) model~\cite{ya}, the scale $\Lambda$ is
\be
\Lambda^4 = e^{-\frac{2\pi}{\alpha_2(M_{Pl})}} \epsilon^{10}
M^3_{SUSY} M_{Pl} \, ,
\label{lambda}
\ee
where $M_{SUSY}$ is the soft SUSY-breaking mass scale,
$\epsilon$ a flavor symmetry breaking parameter
and $\alpha_2(M_{Pl})$ the electroweak gauge coupling strength
at the Planck scale. It is straightforward to verify that
for $M_{SUSY} \sim $ few TeV, $\epsilon = {\cal O}(1)$ and
$\alpha_2(M_{Pl}) = 1/23$, we get $\Lambda \sim 10$ eV.
Since the axion mass is
\be
m \simeq \frac{\Lambda^2}{f_a} \, ,
\label{axmas}
\ee
we see that for $f_a \sim 4\cdot 10^{8}$ GeV we find
$m \sim 10^{-16}$ eV. As we have seen above, these are
roughly the scales most interesting for photon-axion oscillations
in extra-galactic magnetic fields.

It is important to stress that while our axion must
be light {\it it is not light enough} to be quintessence.
Cosmologically, the axion particles with mass $m\sim 10^{-16}$ eV are
relativistic throughout the history of the Universe, and so would
behave like warm dark matter. Because they are weakly coupled,
with $M^{-1} \sim 10^{-12}$ GeV$^{-1}$, they are out of equilibrium
from a very early time. If they are not significantly
produced during reheating after inflation, their abundance can be
harmlessly small. On the other hand, the homogeneous axion background field
$a(t)$ will oscillate around its minimum, with its
energy density scaling as cold dark matter at late times.
Thus one may worry about the cosmological moduli problem
which such fields usually lead to. However in our case this does not
happen because $f_a \sim 10^{8}$ GeV and $m \sim 10^{-16}$ eV.
In the early Universe, the background field will satisfy the slow roll
conditions, and remain frozen until the Hubble scale comes down
to $H \sim 10^{-16}$ eV, when the Universe
cools to the temperature $T_i \sim 100$ keV. At that moment, the
field may start rolling. Its kinetic and potential energy
will rapidly virialize, after which the energy density stored in
it will scale as $\rho \sim \rho_i (T/T_i)^3$. The initial energy
density is determined by the initial displacement of the axion
from its minimum, which is of order $f_a$. Therefore the energy
density will be of order given in Eq.~(\ref{lambda}),
$\Lambda^4 \sim ({\rm few} \times
1 \, {\rm eV})^4$. This would not compete with radiation
until the temperature comes down to $T \sim \rho_i/T_i^3 \sim
10^{-15}$ eV, which means that even if the axion was displaced from
the minimum it would remain tiny for a long time into the future.
Furthermore, while an axionic sector can give rise to both
domain walls and cosmic strings in the early Universe,
because the axion scales in the model we discuss are so low, these
defects may remain negligible well into the future of our
Universe~\cite{defects}.

In summary, we have presented an alternative explanation of the observed
dimming of SNe at large distances. The effect is based on
a quantum mechanical oscillation between the photon field and a hypothetical
axion field in the presence of extra-galactic magnetic fields.
This would result on average
in about a third of the photons emitted by distant SNe oscillating
into axions. This is, roughly, the right amount needed to explain
the supernova observations.
If the average magnetic field is of the order $10^{-9}$ G,
and the average domain size is of order $\sim $ Mpc, one would need an
axion whose coupling to the photon is given by $M \sim 4\cdot 10^{11}$ GeV,
and mass $m\sim 10^{-16}$ eV. With these parameters the
luminosity-distance vs. redshift curve is almost indistinguishable from
the curve of an accelerating Universe with $\Omega_m=0.3,
\Omega_{\Lambda}=0.7$. Since the precise value of the luminosity-distance
for a particular supernova depends on the details of the inter-galactic
magnetic field, we expect more variations in the observed luminosity, and thus
this model may easily incorporate outlying data points.
However, distinguishing this model from the accelerating Universe
paradigm will likely be easier through improving the
bounds on the couplings of ultra-light axions, by understanding the
details of the intergalactic magnetic field, or by a precise independent
determination of the equation of state for the dark energy component,
for example through the DEEP survey~\cite{DEEP}.

\section*{Acknowledgements}

We thank T.~Bhattacharya for explaining to us the proper procedure to
average over the magnetic field, to A.~Albrecht, S.~Dimopoulos,
J.~Erlich, C.~Grojean, S.~Habib, M.~Kaplinghat,
L.~Knox, A.~Linde and R.~Wagoner for useful
discussions, and to G.~Raffelt for comments on the first version of this 
paper. N.K. thanks the members of the T-8 group at Los
Alamos for their hospitality where this work was initiated. C.C.
is an Oppenheimer fellow at the Los Alamos National Laboratory, and is
supported in part by a DOE OJI grant.
C.C. and J.T. are supported by the U.S. Department
of Energy under contract W-7405-ENG-36. N.K. is supported in part
by an NSF grant PHY-9870115.


\end{document}